\documentclass[twocolumn]{aastex62}

\usepackage{graphicx}	
\usepackage{amsmath}	
\usepackage{amssymb}	
\usepackage{soul}

\received{January 1, 2018}
\revised{January 7, 2018}
\accepted{\today}
\submitjournal{ApJL}

\shorttitle{First radio evidence for impulsive heating of the quiet solar corona}
\shortauthors{Mondal et al.}

\begin{document}

\title{First radio evidence for  impulsive heating  contribution to the quiet solar corona}


\correspondingauthor{Surajit Mondal}
\email{surajit@ncra.tifr.res.in}

\author{Surajit Mondal}
\affil{National Centre for Radio Astrophysics, \\
Tata Institute of Fundamental Research, \\
Pune-411007, India}

\author{Divya Oberoi}
\affil{National Centre for Radio Astrophysics, \\
Tata Institute of Fundamental Research, \\
Pune-411007, India}

\author{Atul Mohan}
\affil{National Centre for Radio Astrophysics, \\
Tata Institute of Fundamental Research, \\
Pune-411007, India}
\affil{Rosseland Centre for Solar Physics,\\
Institute of Theoretical Astrophysics,\\
University of Oslo, Postboks 1029 Blindern, N-0315 Oslo, Norway}



\begin{abstract}
{
This Letter explores the relevance of nanoflare-based models for heating the quiet sun corona.
Using meterwave data from the Murchison Widefield Array, we present the first successful detection of impulsive emissions down to flux densities of $\sim$mSFU, about two orders of magnitude weaker than earlier attempts. 
These impulsive emissions have durations $\lesssim 1$ s and are present throughout the quiet solar corona. 
The fractional time occupancy of these impulsive emissions at a given region is $\lesssim 10\%$.
The histograms of these impulsive emissions follow a power-law distribution and show signs of clustering at small timescales.
Our estimate of the energy that must be dumped in the corona to generate these impulsive emissions is consistent with the coronal heating requirements.
Additionally, the statistical properties of these impulsive emissions are very similar to those recently determined for magnetic switchbacks by the Parker Solar Probe (PSP). 
We hope that this work will lead to a renewed interest in relating these weak impulsive emissions to the energy deposited in the corona, the quantity of physical interest from a coronal heating perspective, and explore their relationship with the magnetic switchbacks observed by the PSP.
}
\end{abstract}




\section{Introduction}

The solar corona, or the outermost layer of the solar atmosphere, is at a temperature of about 1 MK, while the photosphere is at a much lower temperature of $\sim$5800 K.
It is now well accepted that the convective motions below the photosphere move the magnetic footpoints randomly building up magnetic stress, which ultimately gets converted to heat. However, the details of how this magnetic energy is converted to heat are not well understood.
There is an increasing realization that an impulsive heating scenario where heat is dumped randomly into the corona might be the dominant mechanism of this energy conversion.
However, other modes of energy conversion also exist and are being studied actively \citep[e.g.][]{klimchuk06}.
 
\citet{klimchuk15} defines nanoflares as small impulsive heating events occurring on small spatial scales without regard to the actual physical mechanism.
In this work, we refer to all events responsible for impulsive heating as nanoflares.
\citet{hudson1991} showed that for nanoflares to be important for coronal heating $\alpha$ must be  $>2$, where $N(E)\propto E^{-\alpha}$ and $N(E)$ is the number of nanoflares with energy $E$.
We refer to this as the Hudson criterion.
\citet{aschwanden00} showed using data from EUV to HXR (spanning the energy range $10^{24}$ to about $10^{32}$ ergs) that $\alpha= 1.79 \pm 0.08$.
This does not satisfy the Hudson criterion, implying that in the energy range where it has been established, the observed flares are not responsible for coronal heating.
However, all methods of the class used by \citet{aschwanden00}, which rely on removing a background, are prone to undercounting flares at low energies due to limitations from sensitivity and resolution of the instruments.
\citet{pauluhn07} followed a different approach, where they tried to find a model of nanoflares to match the statistical properties of the observed light curve.
They showed that the model that best fits the data has $\alpha >2$, meeting the Hudson criterion.
There are several other pieces of observational evidence supporting a nanoflare-based heating scenario, e.g. the high degree of variability observed in active region moss \citep{testa13,testa14}, highly correlated light curves in widely separated filters \citep{viall12,viall17}.
Many studies using radio data \citep[e.g.][etc.]{mercier1997,ramesh13,suresh17} have shown that type I bursts, which are generally associated with active regions, satisfy the Hudson criterion.
A recent detailed multiwavelength spatially resolved study of a weak flaring site associated with a coronal loop finds evidence for episodic impulsive heating \citep{mohan2019}.
A consensus is being slowly reached in the community that the active regions and coronal loops may be heated impulsively. 

In the case of the quiet sun, the answer is unclear. 
While simulations show that steady heating scenarios cannot explain the observed properties of coronal loops, it might still be possible for such heating to operate in the quiet sun \citep{klimchuk10}. 
Some works \citep[e.g.][]{pauluhn07,hahn2014} show that nanoflares may be important for coronal heating, though the final verdict on this is not out yet.
\citet{sharma18a} showed that the energy  radiated in the slowly varying component, dominated by thermal bremsshtrahlung, and the impulsive nonthermal component of the solar meterwave emission, which arises in the corona, are of similar magnitude even during fairly quiet times.
These data, however, were not sufficient for a robust determination of $\alpha$.

We use data from the Murchison Widefield Array \citep[MWA;][]{lonsdale09,tingay13} to investigate the relevance of nanoflare-based heating for the quiet sun. 
The key advantage of using meterwave observations is that the observational signatures of these nonthermal emissions are intrinsically very bright. 
This allows radio observations to probe much weaker energetics than possible with the current generation of instrumentation in EUV and X-rays.
Additionally, the ground-based radio observations also offer a much higher temporal resolution.
While these advantages have long been appreciated, it is only recently that the steady march of technology has enabled radio instrumentation capable of imaging the {\em quiet} sun with sufficient time resolution and imaging fidelity.
In addition, to deal with the data deluge from the modern instruments and make studies of this kind feasible, which require tens of thousands of solar radio images, one needs an unsupervised automated imaging pipeline with a robust performance.
We have recently developed a pipeline that meets these requirements - Automated Imaging Routine for Compact Arrays for the Radio Sun \citep[AIRCARS;][]{mondal19}.


Section 2 describes the observations and the state of the sun on that day. 
The results and a discussion of their implications are presented in Section 3 and Section 4. Section 5 gives the conclusions from this work.

\section{Observations}

We use data from the MWA taken on 2017 November 27.
This day is characterized by a very low level of solar activity\footnote{\url{https://www.solarmonitor.org/?date=20171127}}.
No X-ray flares were reported by Geostationary Operational Environmental Satellite (GOES) in the neighboring two days.
Only one active region (NOAAA 12689) was present on the visible part of the solar disk.
No radio flare was reported on this day.
No other active region was seen by STEREO-A, which was at an angle of $123.5^{\circ}$ with respect to the Sun-Earth line.
The Global Oscillation Network Group (GONG) farside line-of-sight magnetogram also did not reveal any strong magnetic feature.
So, the level of solar activity was very low on the far side of the Sun as well.
Of all the MWA data available, these are among the most suited for exploring the low-level quiet Sun variability.

\begin{figure}
  \centering
    \includegraphics[trim={2cm 1cm 2cm 0.5cm},clip,scale=0.45]{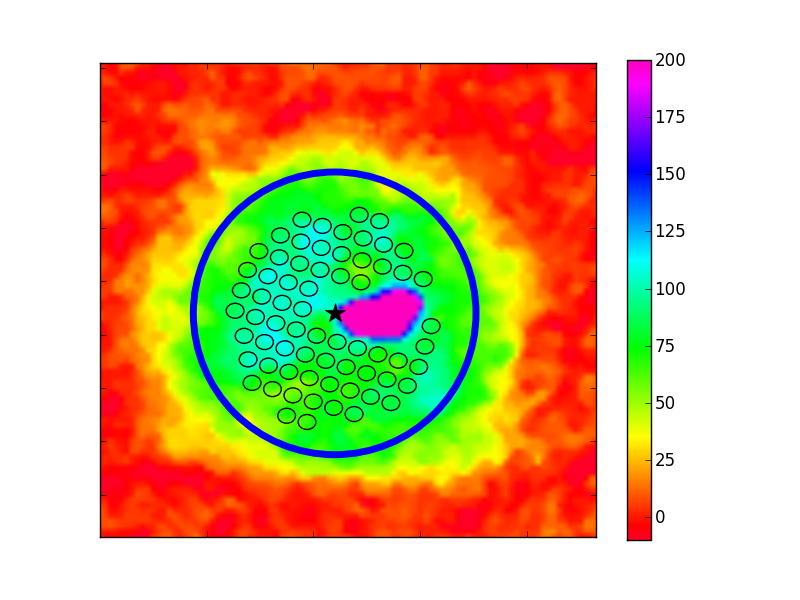}
    \caption{An example solar image at 160 MHz (0.5s, 160 kHz resolution). The blue circle represents the photospheric solar disk. The color scale is in arbitrary units and has been saturated at 200, to highlight the featureless solar disk, apart from the lone active region. The black ellipses indicate the psf-sized regions, the flux density from each of which have been used.
    }
    \label{fig:solar_image}
\end{figure}

On this day, MWA observations were available from 01:30 UT to 03:38 UT.
The observations were done in 12 frequency bands each of 2.56 MHz bandwidth, centered near 80, 89, 98, 108, 120, 132, 145, 161, 179, 196, 217, and 240 MHz.
Of these we have analyzed 70 minutes of data starting from 01:30 UT at four of the frequency bands centerd near 98, 120, 132, and 160 MHz.
Imaging was done using the AIRCARS at a 0.5 s cadence and 160 kHz frequency resolution, using the default parameters. 
This leads to a total of about 33,000 images.
A typical image is shown in Fig. \ref{fig:solar_image}.

\begin{figure*}
    \centering
    \includegraphics[trim={5.2cm 1cm 0cm 1cm},clip,scale=0.5]{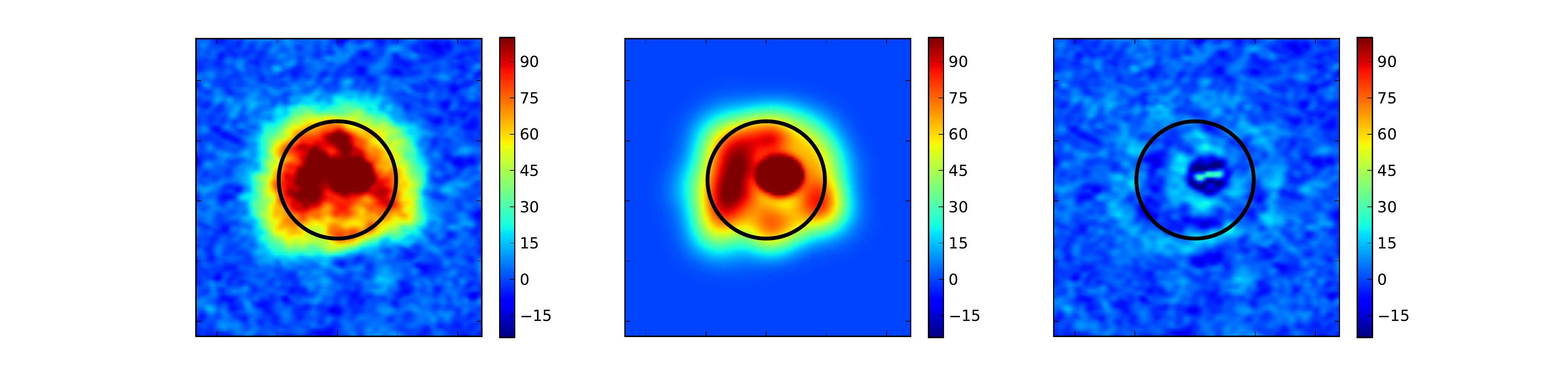}
    \caption{Left panel: an example 120 MHz image made using \textit{Multiscale Clean}; Middle panel: the corresponding model for solar emission produced by the same algorithm (after convolution with the restoring beam to facilitate comparison). Right panel: the residual image or the difference between the calibrated data and the model. The black solid line represents the optical disk of the sun. In the left and the middle panel, the color scale has been saturated to make the faint extended emission visible. The peak values for the left, middle, and right panels are 2487, 2458, and 28, respectively.
    }
    \label{fig:model_image}
\end{figure*}

In order to model the comparatively featureless large angular scale emission of the quiet Sun reliably, AIRCARS uses the \textit{Multiscale Clean} algorithm \citep{cornwell2008}.
This algorithm is tailored to improve the convergence and stability of the conventional Clean when dealing with emission at large angular scales, and a robust implementation of this algorithm is available in the package Common Astronomy Software Applications  \citep{mcmullin07}.
Fig. \ref{fig:model_image} shows an example solar map, model, and the residual generated using \textit{Multiscale Clean}. 
The residual map represents the sum of contributions from the instrumental and sky noise, calibration errors, and deconvolution errors along with all of the unmodeled sky emission.
It is evident that the model is able to adequately capture the much weaker large angular scale emission associated with the quiet sun even in the presence of a much brighter compact nonthermal source.  
The peak unmodeled emission in the residual map is about 88 times weaker than the peak emission in the image and 3 times weaker than the emission from the extended solar disk.

The dynamic ranges (DRs) of the images presented in this work vary significantly with frequency and time. 
The typical DRs at 98, 120, 131, and 160 MHz were 150, 500, 800, and 1200. 
A type I noise storm seemed to be in progress at NOAAA 12689.
In view of the DR limitation, periods of significant activity at the site of the noise storm were not included in this study.
  
\section{Results}

The entire solar image was tiled using point-spread-function (psf) sized patches.
The psf size is a strong function of frequency and remains essentially unchanged across our observations.
The psf major axes at 98, 120, 131, and 160 MHz are 375", 170", 160", and 112", respectively, and the axial ratio is about 1.2.
Data only from regions with signal-to-noise ratio (S/N) $\gtrsim 6$ were used.

\begin{figure}
    \centering
        \includegraphics[trim={3.5cm 1.5cm 1.5cm 1.5cm},clip,scale=0.37]{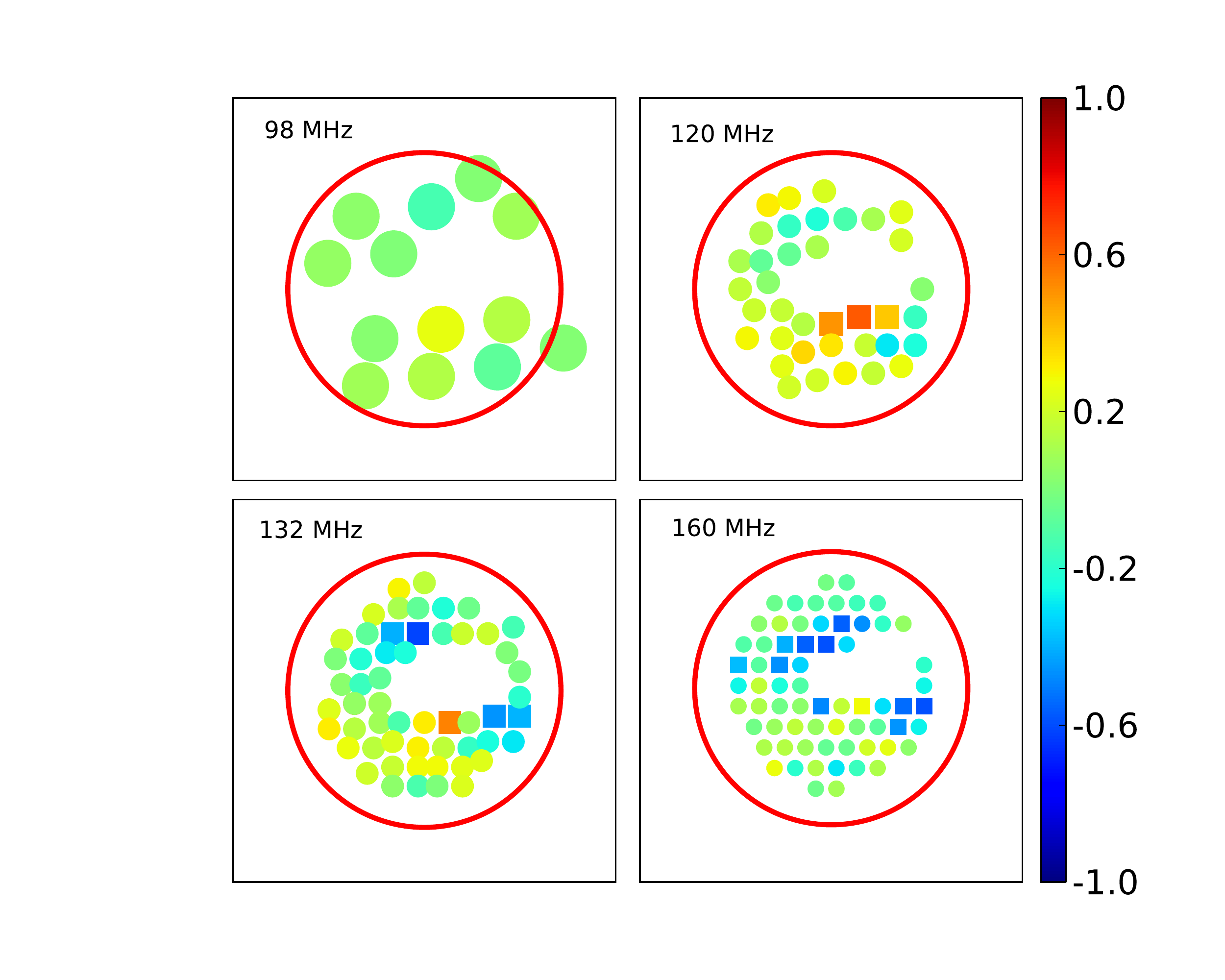}
        \caption{Correlation coefficient as a function of the location of the tile for each of the four frequencies studied here (see the text for details). 
        The red circles show the optical disk of the sun. The squares mark the regions that were excluded from subsequent analysis due to a high correlation coefficient.}
        \label{fig:correlation}
\end{figure}

The flux density time series was extracted for every region of every frequency, and the median flux density was computed.
We denote the median flux density at region $i$ and frequency $\nu$ as $\left<F_{i,\nu}\right>$.
We define $(\Delta F/F)_{i,\nu}=\left(F_{i,\nu}-\left<F_{i,\nu}\right>\right)/\left<F_{i,\nu}\right>$.
As the focus of this study is the quiet Sun, we exclude the regions in the vicinity of the only active region.
Care was taken to assess and avoid any possible contamination to the quiet sun regions used in this study from the intensity fluctuations in the active region.
The flux density time series from the active region was correlated with the corresponding time series from each of the quiet sun regions. 
Figure \ref{fig:correlation} shows the correlation coefficients thus obtained.
For the vast majority of the patches the correlation coefficients lie between $\pm$0.2, implying a lack of evidence for a significant flux leakage.
Exercising an abundance of caution, we have only included patches with correlation coefficients lying between $\pm$0.4.
This leads to the rejection of 0\%, 8\%, 9\% and 15\% of the regions at 98, 120, 132 and 160 MHz respectively.
In order to ensure high S/N, for each region $i$, data points for which $F_{i,\nu}<\left<F_{i,\nu}\right>$ were also excluded.
During quiet times, the solar radio emission is believed to be dominated by the thermal component.
The presence of any nonthermal component can only add to it.
Hence, at quiet sun regions, it is reasonable to expect $\left<F_{i,\nu}\right>$ to be representative of the thermal component.
Not including data below $\left<F_{i,\nu}\right>$ does not bias any investigation of the nonthermal component.
To minimize any contaminating effects due to scattering (which increase as one approaches the limb), while having a sufficient number of data points to work with, we only use regions within 0.8$R_\odot$ except at the lowest frequency.

\subsection{Flux density histogram}

\begin{figure}
\centering
\includegraphics[scale=0.4]{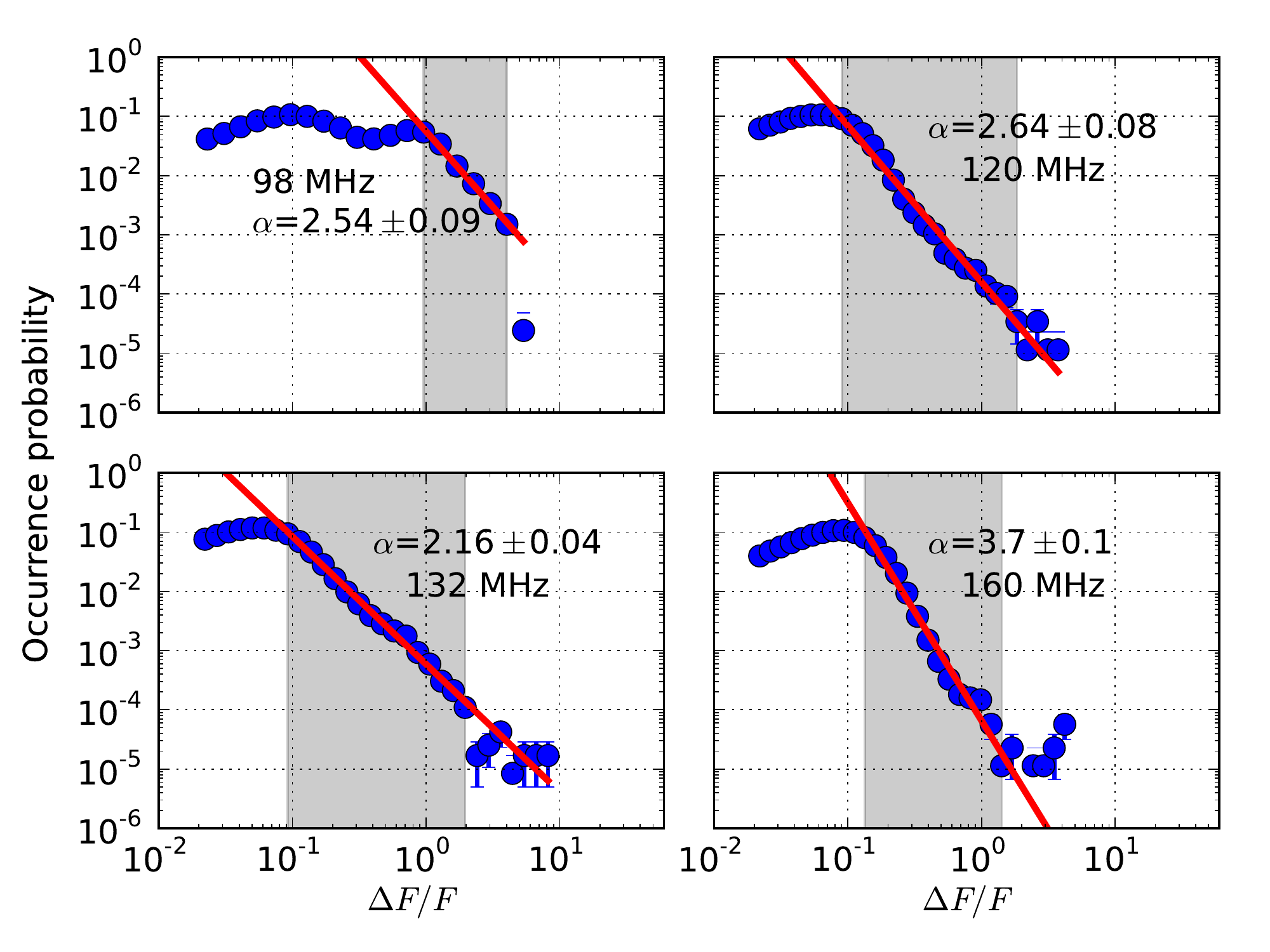}
\caption{Occurrence probability of $\Delta F/F$ at four frequencies. The frequency and fitted value of $\alpha$ are shown in each panel. The gray regions show the area that was used for fitting the power law.}
\label{fig:histograms}
\end{figure}

For every frequency, data satisfying the selection criteria given above are combined, and a histogram of $\Delta F/F$, the occurrence probability, is made (Fig. \ref{fig:histograms}).
The error bars on each data point in the histogram have been obtained assuming Poisson statistics, and are usually too small to be evident in Fig. \ref{fig:histograms}.
The tails of each of these histograms are fit well by a power law, shown by the red curve in the figure.
Some of the data points at high $\Delta F/F$ have been excluded from the fit, due to their large Poisson uncertainties.
The power law spans $\sim$1 order of magnitude along the x-axis and 3--4 along the y-axis.
We find that in all of the cases, $\alpha> 2$ at a significance between 3$\sigma$ and 16$\sigma$.

The energy radiated away (the observed quantity) is related in a nonlinear manner to the energy deposited in the corona by the corresponding event.
Hence, the power-law index of the energy deposition events is expected to be different from that derived here from the observed radiated power. 
However, simulations are starting to hint that the power-law index of the flux density distribution is shallower than that of the energy deposition event distribution \citep{bingert13}.
The results obtained here can only be used as evidence in favor of nanoflare-based coronal heating theories once this is verified by more detailed and extensive analysis. 

The Radio Solar Telescope Network (RSTN) has measured the average noon time solar radio flux at 245 MHz around the days of our observation, characterized by very low levels of activity, to be $\sim$20 SFU (S. White 2020, private communication).
This value is the very close to the flux value observed at a very nearby frequency by \citet{Oberoi2017} during quiet conditions.
It is reasonable to expect that the solar flux densities at other frequencies in the MWA band would also be very similar to those measured by \citet{Oberoi2017}.
This leads to flux density estimates of $\sim$3 and $\sim$6 SFU at 120 and 160 MHz, respectively.
The flux-calibrated images used for this work were generated using the prescription provided by \citet{mohan2017}.
Using this relatively coarse calibration already implies that the typical mean flux density of the regions used in this work are of order 10 mSFU. 
As the weakest impulsive events modeled here lie at $\Delta F/F \sim 0.1$, their flux densities are of order mSFU, making this the weakest detection of nonthermal impulsive features yet.

We use a bootstrapping approach to verify the robustness of these powerlaw fits. 
A thousand realizations, obtained by randomly drawing half the number of data points (with repetition) used in Fig. \ref{fig:histograms}, were generated for each frequency and the best-fit power laws obtained.
A weighted mean of the best-fit power-law indices computed is regarded as the output from the bootstrapping procedure.
The power-law indices thus estimated for 98 MHz, 120 MHz, 131 MHz, and 161 MHz are $2.35 \pm 0.01$, $2.72 \pm 0.01$, $2.155 \pm 0.006$, and $3.55\pm 0.02$ respectively.
All values of $\alpha$ are consistent at $2\sigma$ level with the earlier estimates using the full dataset and are $>2$ at high significance levels, demonstrating the robustness of these fits.

\subsection{Temporal widths of ``events"}

\begin{figure}
\includegraphics[scale=0.42]{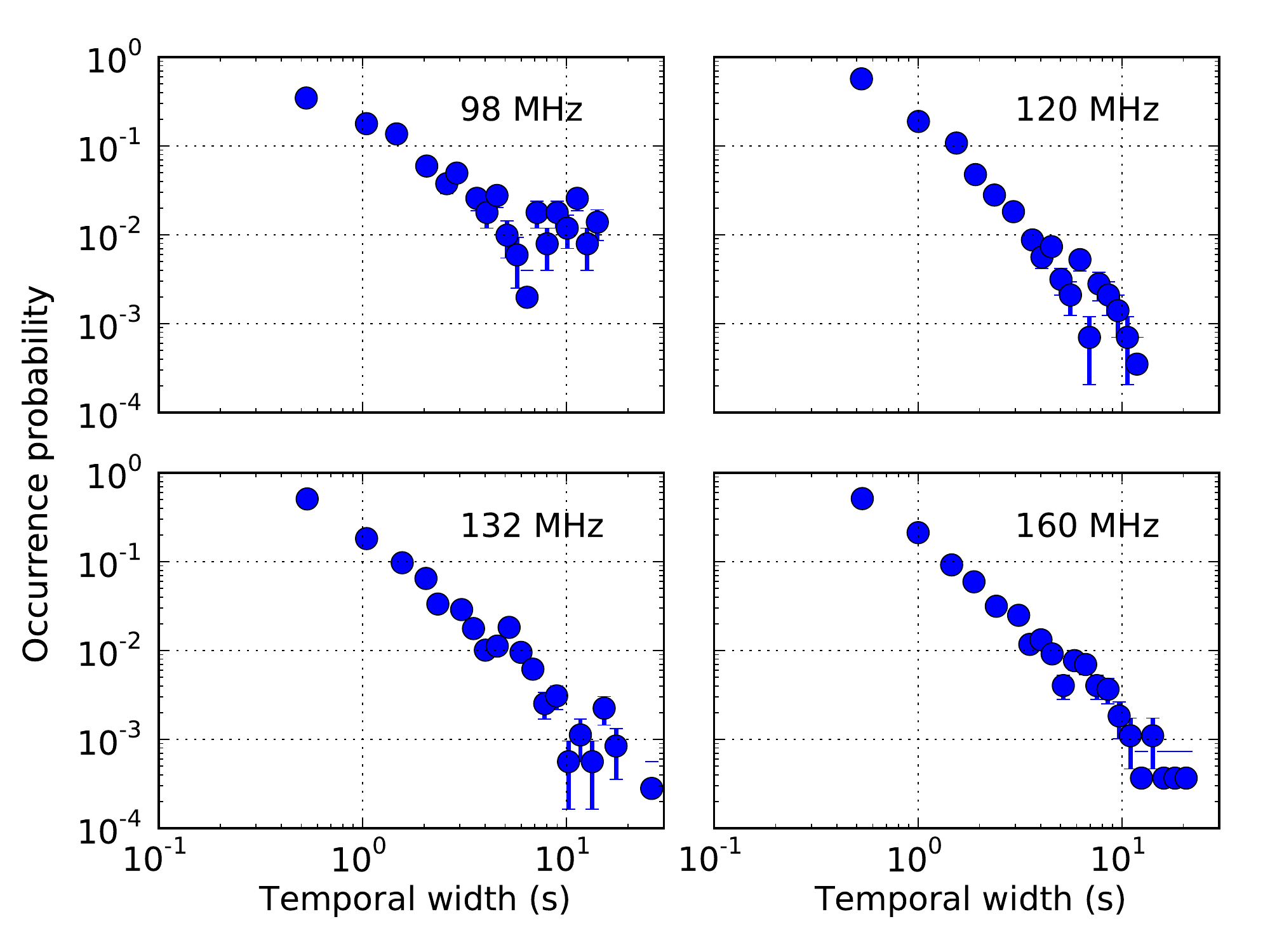}
\caption{Distribution of the temporal widths of the events detected here. It is evident that the vast majority of the events detected here have widths less than or equal to the instrumental temporal resolution of 0.5 s.}
\label{fig:widths_impulsive}
\end{figure}

As the definition of nanoflares adopted here \citep{klimchuk15} requires them to be impulsive emissions, we examine their durations to check if they satisfy this criterion.
An event is defined to be an occurrence of $\Delta F/F$ in the power-law regime, and its duration is defined as the time span for which the $\Delta F/F$ from a region continuously lies above the minimum $\Delta F/F$ to which the power law was fit (Fig. \ref{fig:histograms}).
The observed distribution of durations of these events is shown in Fig. \ref{fig:widths_impulsive} on a log-linear scale.
It is evident that these events are impulsive in nature with durations of the vast majority of them lying close to the instrumental resolution of 0.5 s.
It is noteworthy that at the short duration end, this distribution has a power-law slope close to 2; by the time the duration of the events increases by about an order of magnitude to 5 s, their occurrence rate falls by two orders of magnitude.
The median time duration is $\leq 1$ s at all the frequencies.

This also explains why earlier sensitive studies, though mostly at much higher radio frequencies, looking for exactly such emissions were only able to detect a handful of instances of nonthermal transient brightenings away from active regions \citep{krucker1997,nindos1999}. 
The most sensitive such study that we are aware of is by \citet{nindos1999} using the Very Large Array.
This study included observations at 330 MHz, used snapshot images with a 10 s time resolution, and found one transient brightening away from any active region.
As one averages over a duration an order of magnitude longer than the narrow intrinsic width of the impulsive emission, the signature of this weak emission gets increasingly diluted, dropping below the detection threshold.
Additionally, given the very steep distribution of their flux densities (Fig. \ref{fig:histograms}), the events bright enough to be observable at low time resolutions are too infrequent for many of them to occur in a typical observing span.
We believe that a confluence of these effects lead to the lack of success of earlier efforts.
High imaging dynamic range is another necessary requirement for detecting these faint flux enhancements with a high level of significance.

\subsection{Spatial distribution of ``events"}

\begin{figure}
    \centering
    \includegraphics[trim={2.7cm 0 0 0},clip,scale=0.4]{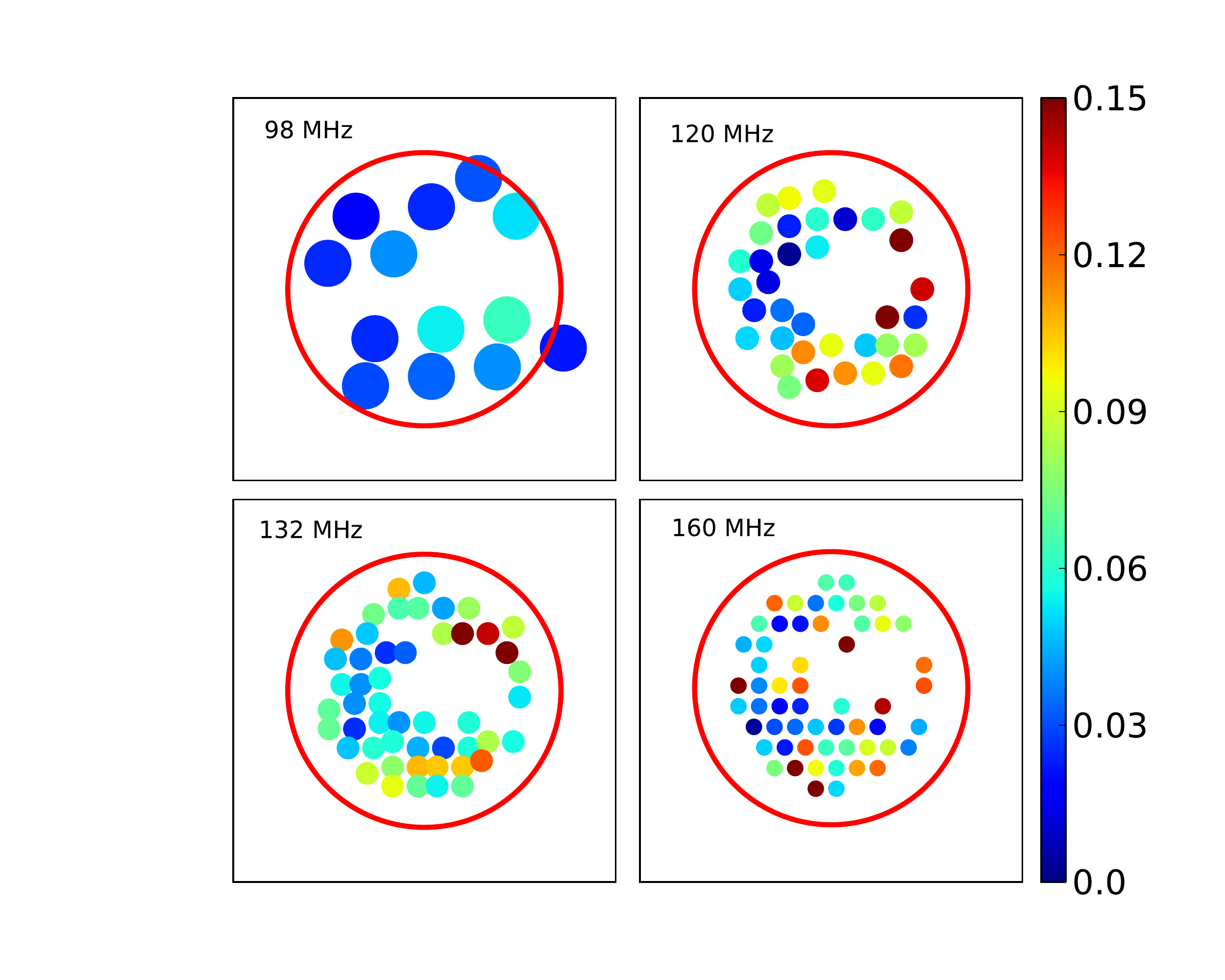}
    \caption{The spatial variation of $\eta$ is shown. The color scale has been saturated at 0.15  so that the non-outlier regions are better visible. 
    The maximum $\eta$ for 98, 120, 132, and 160 MHz are 0.06, 0.29, 0.24, and 0.23, respectively.} 
    \label{fig:spatial_distribution}
\end{figure}

To be relevant for coronal heating, these impulsive emissions need to be ubiquitous in the quiet corona.
To assess this we define $\eta$ to be fraction of time for which the observed flux density for a given region exceeded the minimum $\Delta F/F$ used for the power-law fit.
Figure \ref{fig:spatial_distribution} shows $\eta$ for each of the regions used for this study.
While $\eta$ does show variation across the disk, the
median values of $\eta$ at 98, 120, 132 and 160 MHz are 0.03, 0.07, 0.06 and 0.07.
The minimum values of $\eta$ at 98, 120, 132 and 160 MHz are 0.018, 0.002, 0.025 and 0.003 respectively.
Even the lowest value of $\eta$ corresponds to $\sim$15 events in a given region.
This implies that a significant number such impulsive emissions are present all over the disk.

\subsection{Wait time distribution}
\begin{figure}
\centering
\includegraphics[scale=0.4]{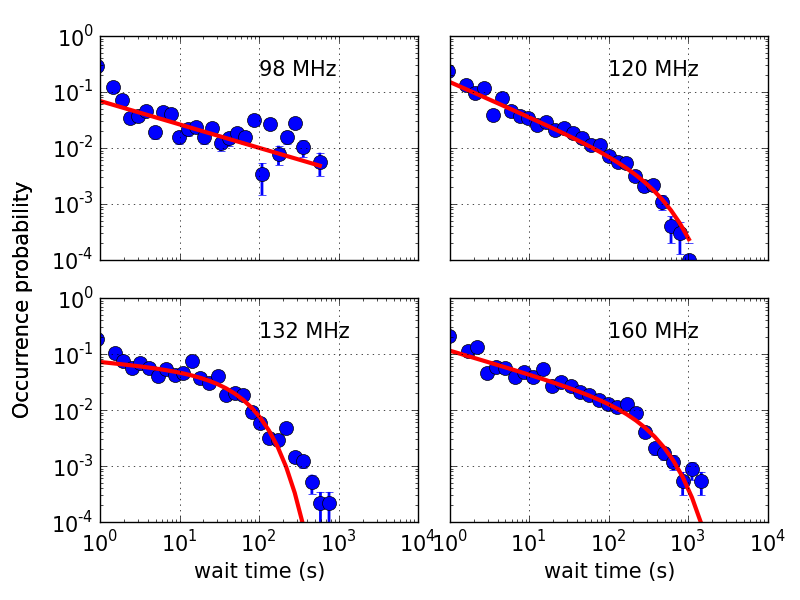}
\caption{Occurrence probability for different wait times at four frequencies. The frequency is shown in each panel.}
\label{fig:wait_time}
\end{figure}

Next we calculate the wait-time distribution of these impulsive events for each frequency.
Given that the MWA time resolution is $0.5$ s, we regard two successive events as distinct if they are separated by at least $1\; s$.
The resultant wait-time distributions are shown in Fig. \ref{fig:wait_time}.
They cannot be described by an exponential distribution, which implies that the these impulsive events are non-Poissonian in nature.
A nonstationary Poisson process of the form used by \citet{aschwanden2010} is also unable to fit these data well. 
These distributions are described well by a product of a power law and exponential model given by $At^{-n}\exp{(-t/t_c)}$, where $A, n$, and $t_c$ are the model parameters and $t$ is the waiting time.
The best-fit models are shown in red in Fig. \ref{fig:wait_time}. 
The power-law behavior of the wait time distributions at small wait times indicate that there is some clustering of these events at small temporal scales.
On the one hand, such a model has been used to model wait-time distributions of X-ray flares in the past \citep{crosby1996}. 
On the other hand, using data spanning much longer durations, \citet{aschwanden2010} have shown that the wait-time distribution of X-ray flares is consistent with an underlying nonstationary Poisson process.
They argue that the inability to model the observed distribution as such a process in earlier works was primarily due to insufficient data.
It is possible that something similar might turn out to be the case in the radio regime as well.
It is instructive to note a few differences though.
The radio impulses being studied here come from energetically much weaker phenomenon, as compared to the ones that are typically associated with even the weakest X-ray flares.
While the X-ray events are sufficiently strong and infrequent that they could be studied using disk-integrated X-ray observations, the radio impulses are so numerous and weak that they tend to blend into a continuum in the dik-integrated flux density.
Studying the latter necessarily requires imaging observations.
In this work, the weakest detected impulsive emissions are limited by the available temporal and angular resolutions.
Hence, it is possible that the intrinsic distribution of these radio impulses might  differ from that observed in X-rays.

\section{Discussion}

\subsection{Implications for coronal heating}

This work presents the first direct observational evidence for the ubiquitous presence of weak impulsive meterwave radio emissions in the quiet solar corona.
The weakest features we detect are about 1 mSFU in strength, about two orders of magnitude weaker than the weakest such emissions reported earlier. 
Impulsive meterwave radio emissions have traditionally been believed to be arising due to magnetic reconnection events. 
Magnetic reconnection leads to the formation of accelerated electron beams that emit at the local plasma frequency and its harmonic as they decay via plasma instabilities. 
The radiative losses due to these plasma emissions are negligible compared to the collisional losses, and the electron beams ultimately get thermalized after transferring its energy to the ambient plasma. 
Hence more energetic beams traverse longer distances before losing their energy. 
These electron beams are generally responsible for the type III bursts with their characteristic narrow time profiles, and rapid spectral drifts spanning large parts of the radio band \citep[see][for a comparatively recent review]{reid2014}. 

A weaker class of solar nonthermal emissions, the dynamic spectra of which are reminiscent of type III bursts, but span much narrower bandwidths, have been documented comparatively recently \citep[e.g.][]{oberoi2011, suresh17}.
Some similarities of these emissions with type I noise storms have also been noted.
The most recent study of bursts that share these characteristics is by
\citet{mohan2019}, who studied an active region transient brightening event associated with a radio noise storm and an X-ray microflare.
Their estimate of the energy of this event was consistent with a microflare, and despite their ability to detect the implusive radio emission over broader bandwidths, they found individual instances of emission to be limited to around 10 MHz. 
They also found the lifetime of the emission to be consistent with the collisional damping time scale.
This led them to suggest a physical picture where the electron beams are weak enough to be collisionally damped.
Two important implications are that (1) these beams are unable to propagate for long distances and (2) they must be produced at the coronal heights from where the emission is observed.
We hypothesize that the emissions reported here are cousins of such emissions, only multiple orders of magnitude weaker.

Building on the physical picture of numerous weak small-scale magnetic reconnections happening throughout the corona proposed by \citet{parker1988}, we propose that
they lead to the formation of accelerated electron beams that emit via plasma emission. 
As mentioned earlier, these weak electron beams thermalize quickly and hence cannot travel far.
Although individually they are energetically weak, their large frequency of occurrence and $\alpha>2$ imply that collectively their contributions can add up to significant amounts.
In about 70 minutes of data, we detect 4748, 24,718, 33,481, and 18,797 events at 98, 120, 132, and 160 MHz, respectively.

It is instructive to attempt an order-of-magnitude estimate of the energy deposited in the corona, despite the intrinsic limitations and uncertainties associated with such an effort.
We do this using the information available from prior work by  \citet{ramesh13}, which provides an estimate of the radiated energy for SFU level emissions, and assuming that it is appropriate to scale it to the kind of events studied here; based on \citet{prasad2004} we use a radio radiative efficiency of $10^{-7}$ for the weak events being considered here; and our own analysis at 132 MHz provides the occurrence frequency of these weak events.
These lead to an estimate of about $\sim10^{26}\ erg\ s^{-1}$, which is comparable to the total coronal heating budget of the quiet corona \citep{sakurai2017}.
Using other radio frequencies also leads to similar estimates.

These constitute evidence of significant energy releases at large coronal heights, implying the presence of a hitherto unaccounted for contribution to the coronal heating budget. 
As instrumentation that can deliver radio images with sufficient dynamic range, and time and frequency resolution becomes available, it will be very interesting to extend this study beyond the present spectral range to look for the presence of weak impulsive emissions at both higher and lower frequencies.

\subsection{Similarities with magnetic switchbacks}

Curiously, the impulsive events studied here share many similarities with magnetic switchbacks, as recently reported in a detailed macroscopic study by \citet{dudok2020} based on Parker Solar Probe (PSP) data.
Swift and omnipresent reversals of magnetic field in the high corona and the interplanetary medium, which otherwise essentially follow the Parker spiral, are referred to as switchbacks.
Their origin has remained elusive, and potentially they have a role to play in heating the solar wind. 
\citet{dudok2020} find that these omnipresent switchbacks do not have a characteristic magnitude (angle by which the magnetic field changes), waiting time, and duration.
Their magnitudes span the entire range from 0--180$^{\circ}$, and the occurrence frequency decreases monotonically with increasing angles.
Their waiting time and duration distributions are remarkably similar.
\citet{chhiber2020} report that the wait-time distribution is modeled well by a power-law + exponential model, which suggests some sort of clustering of the deflections at small temporal scales.

As discussed earlier, the weak impulsive events studied here are also found to be omnipresent, and the distribution of their occurrence probabilities and durations are described well by power laws.
The wait-time distribution is characterized well by an emperical power-law + exponential model.
Magnetic reconnection has also been proposed as a possible origin of magnetic switchbacks \citep{matteini2014}. 
Though the radio data do not span as large a range as presented by \citet{dudok2020}, the similarities between these phenomena are unmistakable, and perhaps suggestive of a possible common cause.

\section{Conclusions}

We present the first detections of ubiquitous weak impulsive radio emissions from the quiet solar corona.
The weakest features detected are $\sim$mSFU in strength, about two orders of magnitude weaker than the weakest such emissions reported earlier.
As small-scale magnetic reconnections are the most likely source of these emissions, their presence constitutes evidence for the ubiquitious presence of a large number of such reconnections that meet the Hudson criterion and are reminiscent of Parker's nanoflares, though at much lower energies.
This is an excellent illustration of how the coherent nature of these emissions enables meterwave radio observations to probe much weaker energetics than currently feasible at EUV or X-ray bands.
A rough estimate of the energies involved suggests that these events could make a significant contribution to the coronal heating budget.
We hope that this work will engender interest in the community to explore the relationship between the observed impulsive radio emissions and the expected energy deposited in the corona, the quantity of true physical interest from a coronal heating perspective.

We find that the weak impulsive events studied here also share many statistical properties with magnetic switchbacks.
Both these phenomena are believed to originate due to magnetic reconnections.
It will hence be very interesting to investigate the detailed relationship between these impulsive events seen by the MWA and switchbacks observed in the solar wind by the PSP.

\section*{Acknowledgements}

This scientific work makes use of the Murchison Radio-astronomy Observatory, operated by the Commonwealth Scientific and Industrial Research Organisation (CSIRO). We acknowledge the Wajarri Yamatji people as the traditional owners of the Observatory site. Support for the operation of the MWA is provided by the Australian Government through the National Collaborative Research Infrastructure
Strategy (NCRIS), under a contract to Curtin University administered by Astronomy Australia Limited. We acknowledge the Pawsey Supercomputing Centre, which is supported by the Western Australian and Australian Governments. 
We thank Stephen White (AFRL) for providing the solar flux density measured by RSTN for a few days near our observation.
S.M. acknowledges Prasun Dutta (IIT-BHU), Dipanjan Mitra (NCRA-TIFR), Subhashis Roy (NCRA-TIFR), and Hardy Peter (MPS) for useful discussions.  The authors acknowledge support of the Department of Atomic Energy, Government of India, under the project No. 12-R\&D-TFR-5.02-0700.
We thank the reviewer for constructive feedback that has helped us improve the presentation and robustness of this work.
We thank the developers of Python 2.7\footnote{See
https://docs.python.org/2/index.html.} and the various associated packages, especially Matplotlib\footnote{See http://matplotlib.org/.}, Astropy,\footnote{See http://docs.astropy.org/en/stable/.} and NumPy\footnote{See https://docs.scipy.org/doc/.}. This research has made use of NASA's Astrophysics Data System.




\facility{Murchison Widefield Array, SDO (AIA), GOES}
\bibliography{references} 




\end{document}